\documentclass[a4paper,12pt]{article}
\usepackage{graphicx}
\begin{document}
\title{Resonant capture of multiple planet systems under dissipation and stable orbital configurations}
\author{George Voyatzis \\ Department of Physics, Aristotle University of Thessaloniki, Greece, \\ voyatzis@auth.gr} 

\maketitle

\abstract{
Migration of planetary systems caused by the action of dissipative forces may lead the planets to be trapped in a resonance. In this work we study the conditions and the dynamics of such resonant trapping. Particularly, we are interested in finding out whether resonant capture ends up in a long-term stable planetary configuration. For two planet systems we associate the evolution of migration with the existence of families of periodic orbits in the phase space of the three-body problem. The family of circular periodic orbits exhibits a gap at the 2:1 resonance and an instability and bifurcation at the 3:1 resonance. These properties explain the high probability of 2:1 and 3:1 resonant capture at low eccentricities. Furthermore, we  study the resonant capture of three-planet systems. We show that such a resonant capture is possible and can occur under particular conditions. Then, from the migration path of the system, stable three-planet configurations, either symmetric or asymmetric, can be determined.     
} 
\section{Introduction} 
\label{intro}
The study of long-term stability of planetary systems involves numerical simulations of $N$-body models consisting of a large body  of mass $m_0$ (the star) and $N-1$ planets $P_i$ of masses $m_i<<m_0$, $i=1,..,N-1$. All bodies are affected by their mutual gravitational interactions and if $N$ is large then we have to face a complex system with collision singularities. In such systems choreographical solutions can be found by assuming particular central configurations \cite{Terracini}. For real planetary systems it is a challenge  to obtain stable solutions along which planets avoid collisions, but this is not feasible starting with a system of many planets. If $N$ is small, then we can model the system with a near integrable Hamiltonian of few degrees of freedom, where stability issues are also important. For the simplest case, $N=3$, we have the classical three body problem (TBP), which has been studied widely and is well known for its complex dynamics. Nevertheless, particular methods applied to the TBP and for two-planet systems, can be extended in order to study the dynamics of systems with three or more planets.   

New interesting questions about the evolution of planetary systems have arisen in the last twenty years after the discovery of many exo-solar systems with many planets. Presently, about $500$ multiple planet systems have been confirmed but, generally, observational data are not sufficient for an accurate determination of the orbital elements of the planets i.e. the semimajor axes, $a_i$, eccentricities, $e_i$, angles of apside, $\varpi_i$, and the location of the planets on their ellipse given e.g. by the mean anomaly $M_i$. In the following the index $i$ starts counting from the inner planet to the outer one.  It is reasonable to assume that multiple planet systems are located at stable configurations that guarantee the long-term stability of planetary orbits. Considering a two-planet system and the TBP model, long term stability generally requires regular trajectories winding invariant tori in phase space. On the other hand chaotic orbits may lead the planets to close encounters, which destabilize the system causing collisions or escapes. Various Hill's type stability criteria have been proposed, which ensure that the planets cannot suffer close encounters (see e.g. \cite{Veras13}). In many cases of real systems, Hill's criteria are not established and stability is offered by other mechanisms. An important dynamical mechanism is based on the mean motion resonances (MMR) i.e. when the planetary periods have almost rational ratio, $\frac{T_2}{T_1}\approx \frac{p}{q}$, $p$, $q$ are integers \cite{Micht08}. Thus, the observation of many resonant exosolar systems may be explained due to this stability mechanism.          

\begin{figure}
\centering
\includegraphics[width=0.85\textwidth]{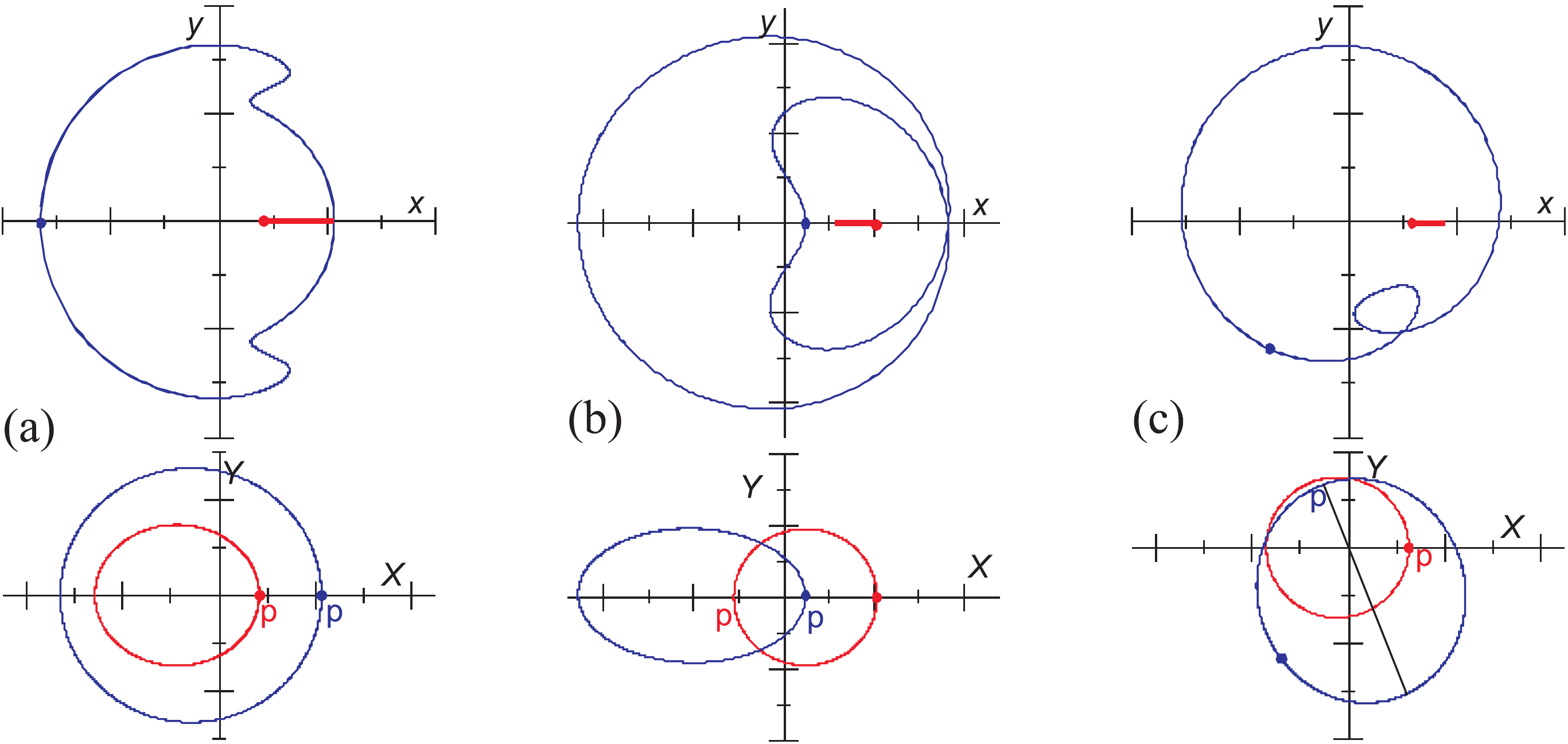} 
\caption{Some samples of 2:1 resonant periodic orbits in the rotating (top) and inertial frame (bottom). Red and blue color indicates the orbit of the inner planet and the outer planet, respectively. In the rotating frame the inner planet oscillates periodically on the $Ox$ axis. Dots indicate the initial planetary position and ``p'' the periastron of (a) a symmetric periodic orbit with aligned periastra (the lines of apsides coincide with the $Ox$ axis) (b) a symmetric periodic orbits with anti-aligned periastra  (c) an asymmetric periodic orbit, the lines of apsides form an angle $\Delta\varpi\neq 0$ or $\pi$ (see also section \ref{Section32}).}
\label{Fig_Porbs}       
\end{figure}

From a dynamical point of view, resonances are indicated by periodic orbits of the particular model, which are called also {\em exact MMR resonances}, in order to distinguish them from the orbits that simply satisfy the rational period ratio. Particularly, for the planar general TBP of planetary type, we define a non-uniformly rotating frame $Oxy$, whose $x$-axis is the line $star - P_1$ and the origin $O$ is located at the center of mass of the two bodies. Thus the position of $P_1$ is determined by the coordinate $x_1$ and $P_2$ by $(x_2,y_2)$. If at $t=0$ the inner planet is at an apside, e.g. at $x_1(0)=x_{10}$  with $\dot x_1(0)=0$ and $\ddot x_1(0)>0$, then a periodic orbit is defined by the conditions 
\begin{equation}
x_2(0)=x_2(T),\quad y_2(0)=y_2(T),\quad \dot{x}_2(0)=\dot{x}_2(T),\quad \dot{y}_2(0)=\dot{y}_2(T),
\label{EqPCon}
\end{equation}
where $T$ is the time (period) for $P_1$ to be found again in its apside with $\ddot x_1(0)>0$ and certainly, by taken into account the energy integral, at the same position $x_{10}$. The period $T$ may corresponds to $k$ revolutions of the inner planet; $k$ is called the multiplicity of the orbit. System (\ref{EqPCon}) is solved numerically by using differential corrections starting close to a known periodic orbit with $x_1(0)=x_{10}$ and seeking a solution for $x_1(0)=x_{10}+\delta$, where $\delta$ is a small deviation. 
In the case where $y_{20}=\dot{x}_{20}=0$ we obtain a periodic orbit which is {\em symmetric} with respect to the $Ox$ axis. Otherwise, a periodic orbit is called {\em asymmetric}. In Fig. \ref{Fig_Porbs}, we present some periodic orbits in the rotating and inertial frame. We note that in the inertial frame the orbits are almost Keplerian ellipses which rotate slowly. Periodic orbits are studied with respect to their linear stability and are classified as stable or unstable. For more details see \cite{Hadjidem06b,Hadjidem06}. 

By continuing the variation of  $x_{10}$ we obtain a monoparametric set of solutions or, equivalently, a {\em family} of periodic orbits. Families are classified as {\em circular} or {\em elliptic}. A circular family consist of of almost circular planetary orbits and the ratio $\frac{T_2}{T_1}$ varies along the family. Along a family of elliptic orbits it is $\frac{T_2}{T_1}\approx \frac{p}{q}=$const and, thus, elliptic families are resonant. The orbits presented in Fig. \ref{Fig_Porbs}  are 2:1 resonant. Generally, the continuation process starts from the known periodic orbits of the circular restricted TBP \cite{Hadjidem75} or the elliptic one \cite{GV09,Antoniadou11}, where the mass $m_1$ (or $m_2$), which is initially zero, is used as the continuation parameter.     

\begin{figure}
\centering
\resizebox{0.7\textwidth}{!}{\includegraphics{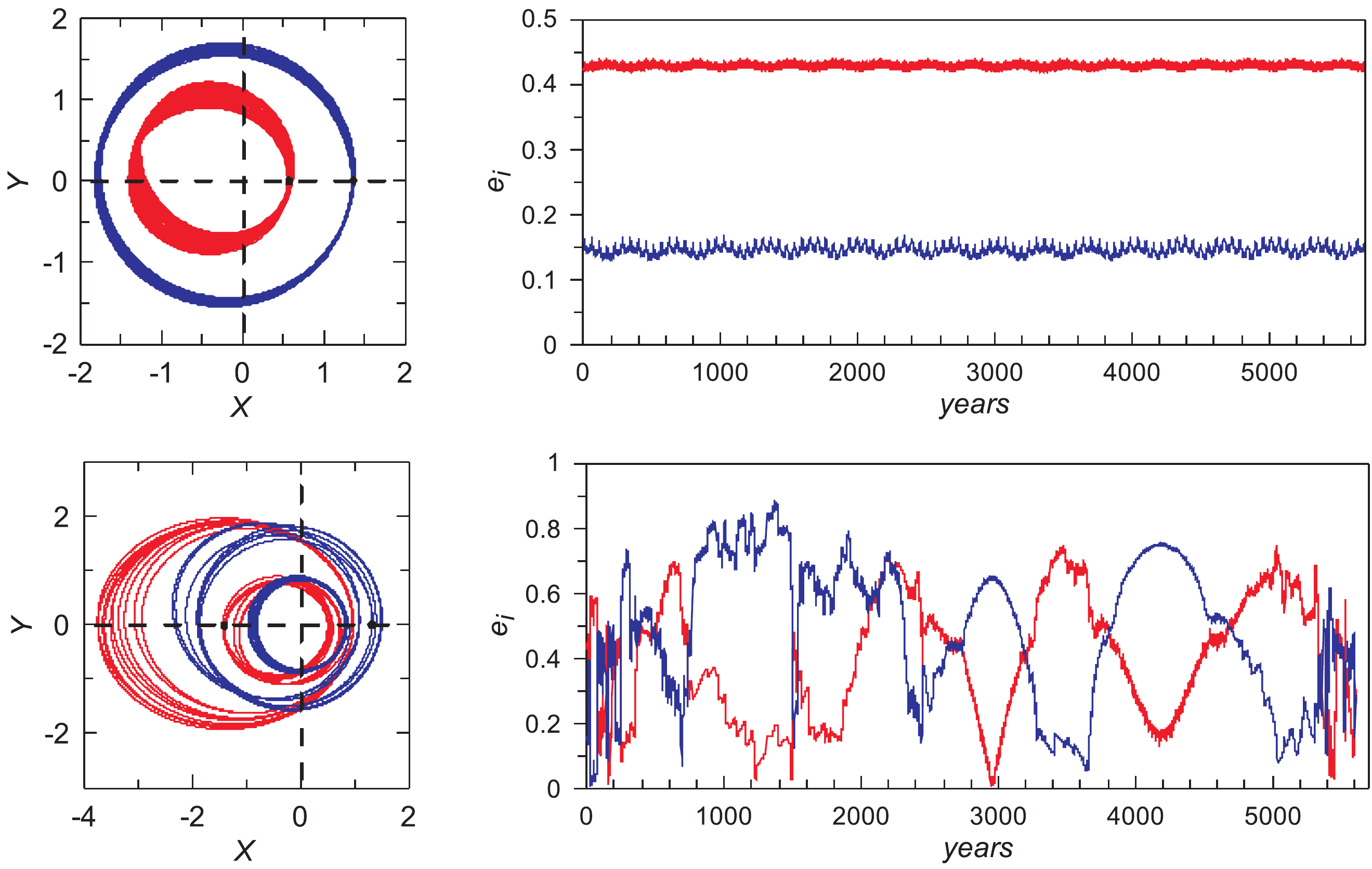}} 
\caption{The evolution of the orbits and the eccentricities of a two-planet system (similar to HD 82943) with masses $m_1\approx m_2\approx 0.004$ and initial conditions $a_2/a_1=1.595$, $e_1=0.425$, $e_2=0.16$ in the symmetric configurations $\varpi_1=\varpi_2=0^\circ$ and $M_1=M_2=0^\circ$ (top) and $M_1=180^\circ$, $M_2=0^\circ$ (bottom).}
\label{figOrdCha}       
\end{figure}

Resonant families may bifurcate from circular periodic orbits but they generally extend up to very eccentric planetary orbits, which may cross each other (e.g. Fig. \ref{Fig_Porbs}b,c). However, the resonance can offer a phase protection mechanism that prevents the planets from coming close to each other. Additionally, if a periodic orbit is linearly stable then orbits which are located in its neighborhood are stable for long-term evolution (assuming that possible Arnold diffusion is not apparent in reasonable time spans). Therefore, in a particular resonance it is important to determine the initial planetary phases (planetary configurations) that correspond to a stable evolution. In Fig. \ref{figOrdCha} we consider the evolution of the orbits and the eccentricities of the two planets, for parameters related to the exosolar system HD 82943 (see e.g. catalog {\em exoplanets.eu}). In the top panel, the system starts close to a stable periodic orbit and we get a regular evolution. In this configuration both planets are initially located at their periastron ($M_1=M_2=0^\circ$) and their orbits are aligned ($\varpi_2-\varpi_1=0^\circ$).  In the bottom panel, we take the same initial conditions except that we initially place the inner planet at its apastron ($M_2=180^\circ$). Now we get a very unstable configuration and the system is destabilized.    

The permanent trapping of a planetary system in resonance seems possible when dissipative forces act in the system \cite{Beauge93}. Such dissipative forces act in the early stage of a planetary system when the protoplanetary disk of gas and dust causes drag and torques to the planetary motion. This has as a consequence the inward migration of the planets from their initial region of formation \cite{Crida07,Chambers09}. Numerical simulations in two-planet systems showed that, during migration, planets can be trapped in resonance \cite{Lee02,Papaloiz03}. Particularly it has been shown that after the resonance capture the system evolves in such a way that it follows particular paths in phase space which consist of families of centers of librations or, equivalently families of stable periodic orbits \cite{Beauge05,Hadjidem10}. Notably, if non-planar motion is considered, an increase of the inclinations is observed at vertically critical periodic orbits \cite{GV14}. After the disk has vanished and the dissipative forces disappear, the planetary system may end up in a stable configuration and continue to evolve regularly. 

In the present study we address the dynamics of migration and resonant capture in two- and three-planet systems. Our model is presented in section 2. It takes into account the gravitational interactions and a Stoke's dissipative force. In section 3, we study the migration of two-planet systems and the conditions for resonance trapping. In section 4, we consider migration of three-planet system and study the stable planetary configurations, which result after the resonant capture. Particular results for the 1:2:4  resonance are presented.  Finally, we conclude in section 5.         

\section{The model} 
We consider the planar $N$-body problem of planetary type. We have a central star of mass $m_0$ and $N-1$ planets of mass $m_i<<m_0$, $i=1,..,N-1$. In a barycentric inertial frame the position of the bodies is given by $\mathbf{r}_i(t)=(x_i(t),y_i(t))$. Assuming that besides the gravitational force between the planets an external dissipative force $\mathbf{F_d}$ may also act on the planets and the equation of motion are
\begin{equation}
m_i \mathbf{\ddot r_i}=\nabla_i U + \mathbf{F_d}_{,i}
\label{EqDeq}
\end{equation}
where $U=U(r_{ij})$ is the potential function of gravitational forces 
$$
U=\sum_{0\leq i <j <N} \sum \frac{m_i m_j}{r_{ij}},\quad r_{ij}=||\mathbf{r}_j-\mathbf{r}_i||. 
$$
Considering that the gas and the dust in the protoplanetary disk both move in circular orbits around the star we can assume as a dissipative force a Stoke's like drag proportional to the relative velocity between the planet and the disk, which is of the form \cite{Beauge93}      
\begin{equation}
\mathbf{F_d}_{,i}=-c_i (\mathbf{v}_i-\alpha_i \mathbf{v}_{c,i}),
\label{EqStokes}
\end{equation}
where $\mathbf{v}_i=\mathbf{\dot{r}}_i$ is the velocity of the planet $i$ and $\mathbf{v}_{c,i}=(m_0/r_i)^{1/2} \mathbf{e_\theta}$ is the velocity of the circular orbit of radius $r_i$ in the two body (star-planet) approximation. The factors $c_i>0$ depend on physical properties of the gas and the size of the planet and are assumed to be constant in the range  $10^{-5}<c<10^{-10}$. The parameters $\alpha_i$ are introduced in order to include the effect of a possible radial pressure gradient in the disk and are assumed to be positive and a bit less than 1. Finally, the equations of motion on the $xy$ plane of the planet $P_i$ are written as
\begin{equation}
\begin{array}{c}
\displaystyle{\ddot x_i=-m_0 \frac{x_i - x_0}{r_{0i}^3}-\sum_{j=1, j\neq i}^{N-1} m_j \frac{x_i-x_j}{r_{ij}^3} - C_i \left (\dot{x}_i+\alpha_i \frac{y_i}{r_{i0}^{3/2}} \right )} \\
\displaystyle{\ddot y_i=-m_0 \frac{y_i - y_0}{r_{0i}^3}-\sum_{j=1, j\neq i}^{N-1} m_j \frac{y_i-y_j}{r_{ij}^3} - C_i \left (\dot{y}_i-\alpha_i \frac{x_i}{r_{i0}^{3/2}} \right ) }            
\end{array},
\label{EqDeqs}
\end{equation}
where $C_i=c_i/m_i$ and the masses are normalized such that $\sum_{i=0}^{N-1} m_i=1$.

In simulations we may assume that some planets are not affected by the dissipation, so in these cases we set $C_i=0$. The divergence of the vector field of system (\ref{EqDeqs}) is 
$$
\textnormal{div} \mathbf{f} = -2\sum_{i=1}^{N-1}C_i <0,
$$ 
thus, the system is dissipative. It has been shown that for an one-planet system and in a first order approximation in eccentricity, such dissipation causes the exponential decrease of the semimajor axis and the eccentricity of the planet \cite{Gomes95,Beauge05},
\begin{equation}
a(t)=a_0 \exp (-A\,t),\quad\quad e(t)=e_0 \exp (-E\,t),
\label{EqExp}
\end{equation}
where 
\begin{equation}
A=2 C (1-\alpha) ,\quad\quad E=C\,\alpha.
\label{EqExpCoef}
\end{equation}      

\begin{figure}
\centering
\resizebox{0.85\textwidth}{!}{\includegraphics{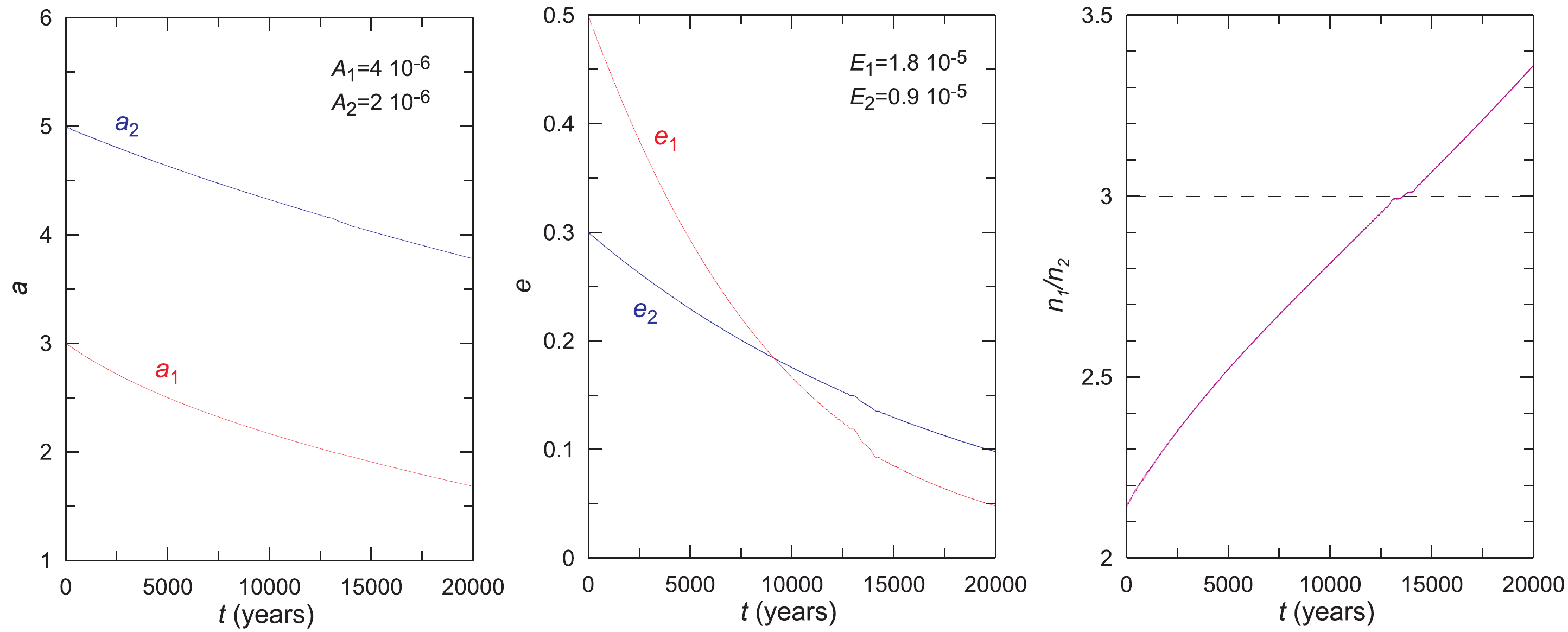}} 
\caption{Evolution of orbital elements of a two-planet system ($m_1=m_2=10^{-4}$) with dissipation, $C_1=2\,10^{-5}$, $C_2=10^{-5}$, $\alpha_1=\alpha_2=0.9$.}
\label{figSMig2P}       
\end{figure}   

\begin{figure}
\centering
\resizebox{0.85\textwidth}{!}{\includegraphics{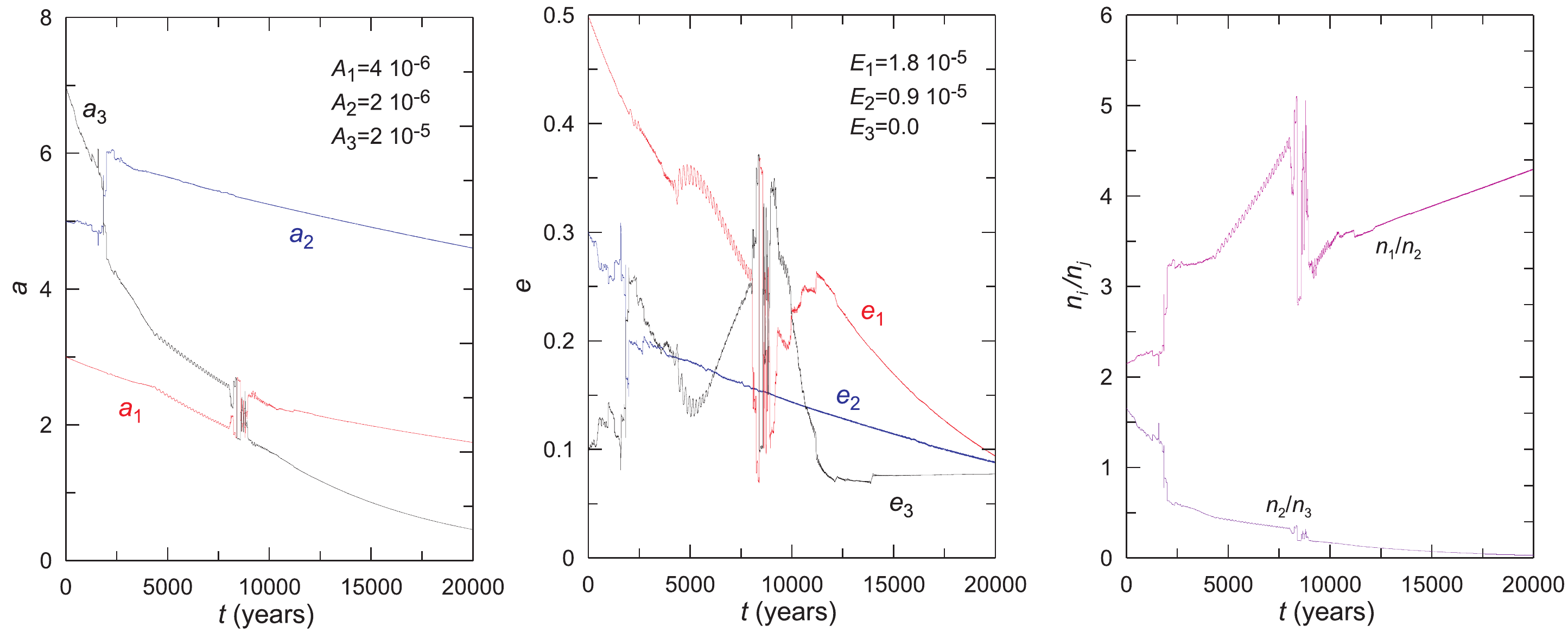}} 
\caption{Evolution of orbital elements of a three-planet system ($m_i=10^{-4}$, $i=1,2,3$) with dissipation, $C_1=2\,10^{-5}$, $C_2=C_3=10^{-5}$, $\alpha_1=\alpha_2=0.9$, $\alpha_3=0$.}
\label{figSMig3P}       
\end{figure}
 
As the semimajor axes vary during migration, the same holds for the Keplerian periods $T_i$ of the planets, which are proportional to $a_i^{3/2}$. Thus, the ratios of $T_i$ between the planets or, equivalently, the ratios of the mean motions $n_i=2\pi/T_i$ vary along the migration. In Fig. \ref{figSMig2P}, we present the variation of semimajor axes, eccentricities and the mean motion ratio of two planets, which interact mutually with gravitation too and are affected by the external dissipative force (\ref{EqStokes}). We obtain an exponential decrease of both semimajor axes and eccentricities. The mean motion ratio increases but when it passes by the value $n_1/n_2=3$ (i.e. the 3:1 resonance) it shows a short plateau, i.e. a temporal resonant capture. In Fig. \ref{figSMig3P}, we present the migration of a three-planet system. After 2Ky we obtain a strong gravitational interaction between the planets $P_2$ and $P_3$. A second strong interaction appears at about 8Ky between $P_3$ and $P_1$. Certainly, the previous strong gravitational interactions affect the evolution of eccentricities and mean motion ratios but, finally, the system seems to enter a regular smooth migrating evolution.  

Apart from dissipative forces of the form (\ref{EqStokes}), other types of dissipation have been used, e.g. two-body tidal interactions \cite{Sylvio03,Delisle14}. Regardless of the type of dissipative forces, the evolution of planetary systems seems qualitatively the same.
 

\section{Migration of two-planet systems}
\subsection{Resonance capture}
For a system of two planets, according to the Kepler's third law and in the absence of mutual gravitational interactions between planets, it is $(a_2/a_1)=(n_1/n_2)^{2/3}$. A necessary condition for the system to be captured in a particular resonance $p:q$, i.e. $n_1/n_2\approx p/q$, is to meet during migration the semimajor axis ratio 
\begin{equation}
(a_2/a_1)\approx (p/q)^{2/3}.
\label{EqRCcon}
\end{equation}
In numerical simulations we set initially the system in a position with $(a_2/a_1)>(p/q)^{2/3}$ and apply the dissipative force (\ref{EqStokes}) only to the outer planet $P_2$.  In this case we expect that the semimajor axis $a_1$ will remain almost constant while $a_2$ will decrease due to the dissipative force. Thus, the ratio $a_2/a_1$, or equivalently the ratio $n_1/n_2$, decreases too and the condition (\ref{EqRCcon}) should be fulfilled after some time of evolution. 

\begin{figure}
\centering
$\begin{array}{ccc}
\resizebox{0.45\textwidth}{!}{\includegraphics{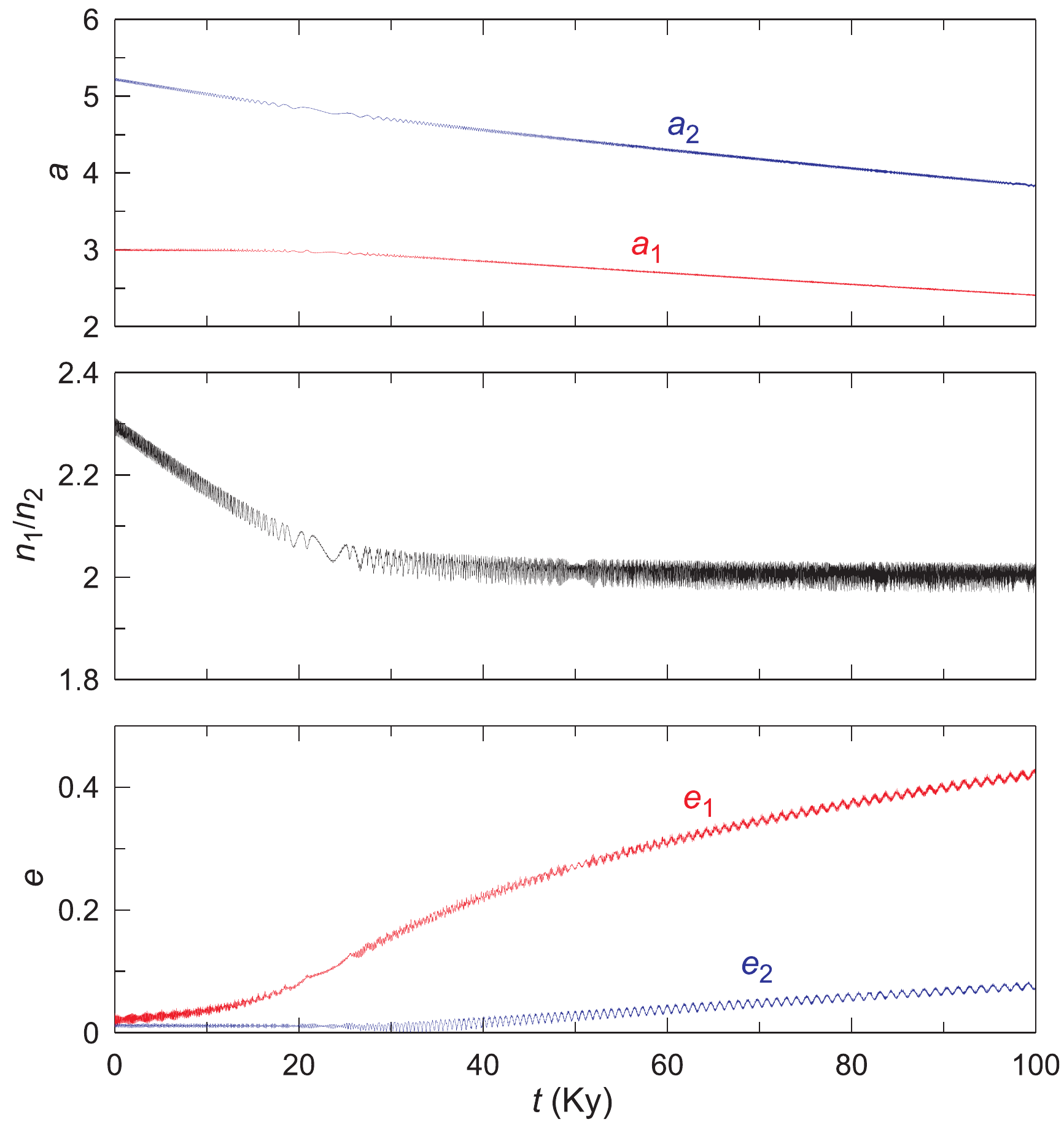}}&\;\;\;\;&\resizebox{0.45\textwidth}{!}{\includegraphics{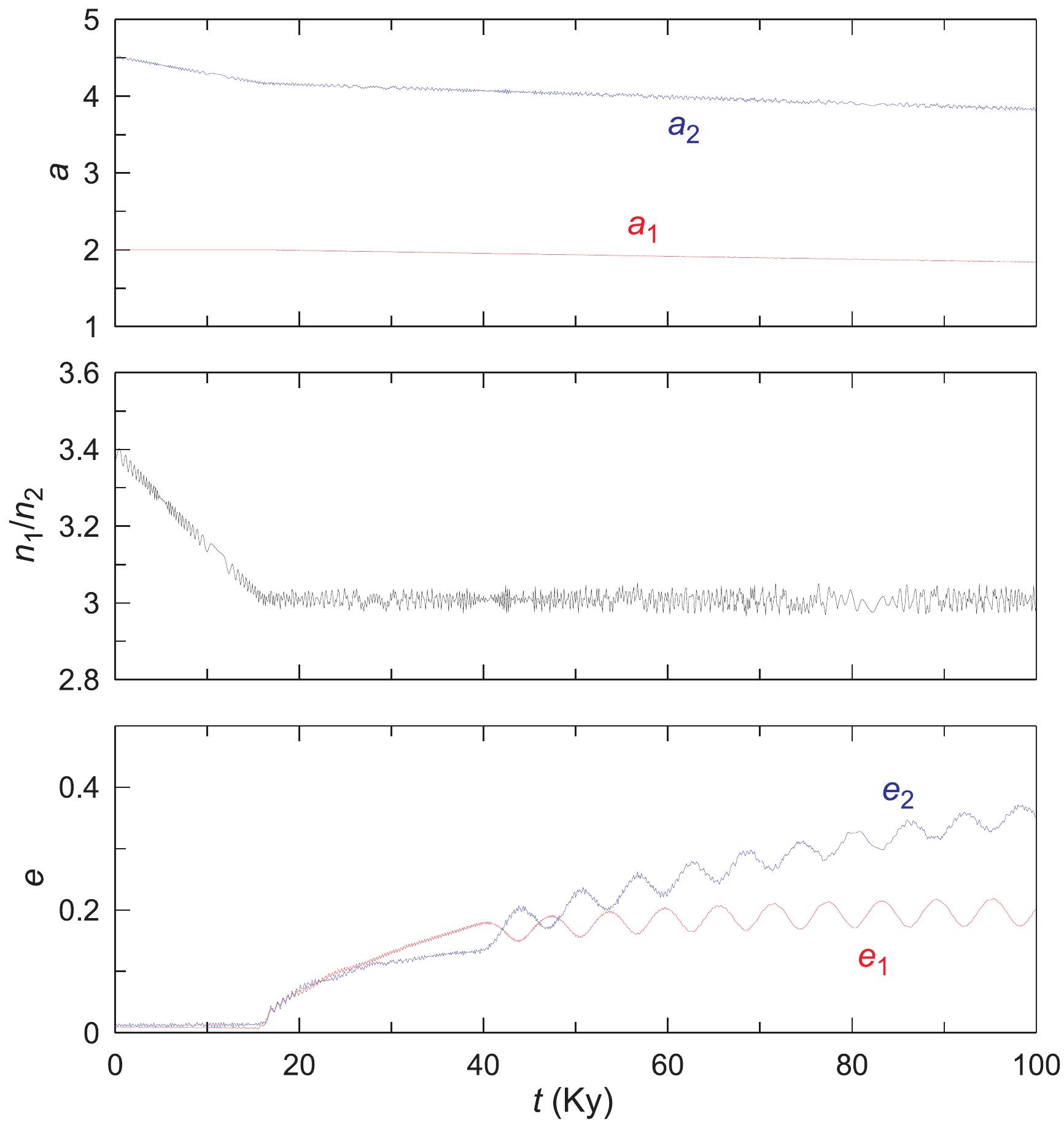}}\\
\textnormal{(a)}&\;\;\;\; & \textnormal{(b)}\\
\end{array}
$
\caption{Evolution of the semimajor axis (top), the mean motion ratio (middle) and eccentricities (bottom) for a two-planet system (a) 2:1 resonance trapping, $m_1=0.001$, $m_2=0.005$, $C_2=10^{-6}$, $\alpha_2=0.7$   (b) 3:1 resonance trapping, $m_1=0.001$, $m_2=0.0005$, $C_2=10^{-6}$, $\alpha_2=0.6$.}
\label{figRcup2131}       
\end{figure}

In Fig. \ref{figRcup2131}, we present two typical examples of migration and capture in resonances 2:1 and 3:1 (left and right panels, respectively). The initial eccentricities of the planets are almost zero (circular orbits). In both cases, the semimajor axis, $a_1$, is initially constant, while $a_2$ decreases. When we obtain the condition (\ref{EqRCcon}), the outer planet seems to drag the inner one to an orbit with smaller and smaller semimajor axis. However, the ratio $a_2/a_1$ remains almost constant and the capture is achieved. After the resonant capture the eccentricities increase. Their rate of increment and the final value that they reach depend on the parameter $\alpha$, which is related with the eccentricity damping parameter $E$ according to Eq. (\ref{EqExpCoef}). 
          
During the trapping in $3:1$ resonance, we should note the change in the evolution of eccentricities for $t>40$Ky (see Fig. \ref{figRcup2131}b). The eccentricities start to show oscillations and the average rate of their increment changes. This change in the evolution is caused by a change in the migration path as it will be explained in next subsection. 

\begin{figure}
\centering
\resizebox{0.7\columnwidth}{!}{\includegraphics{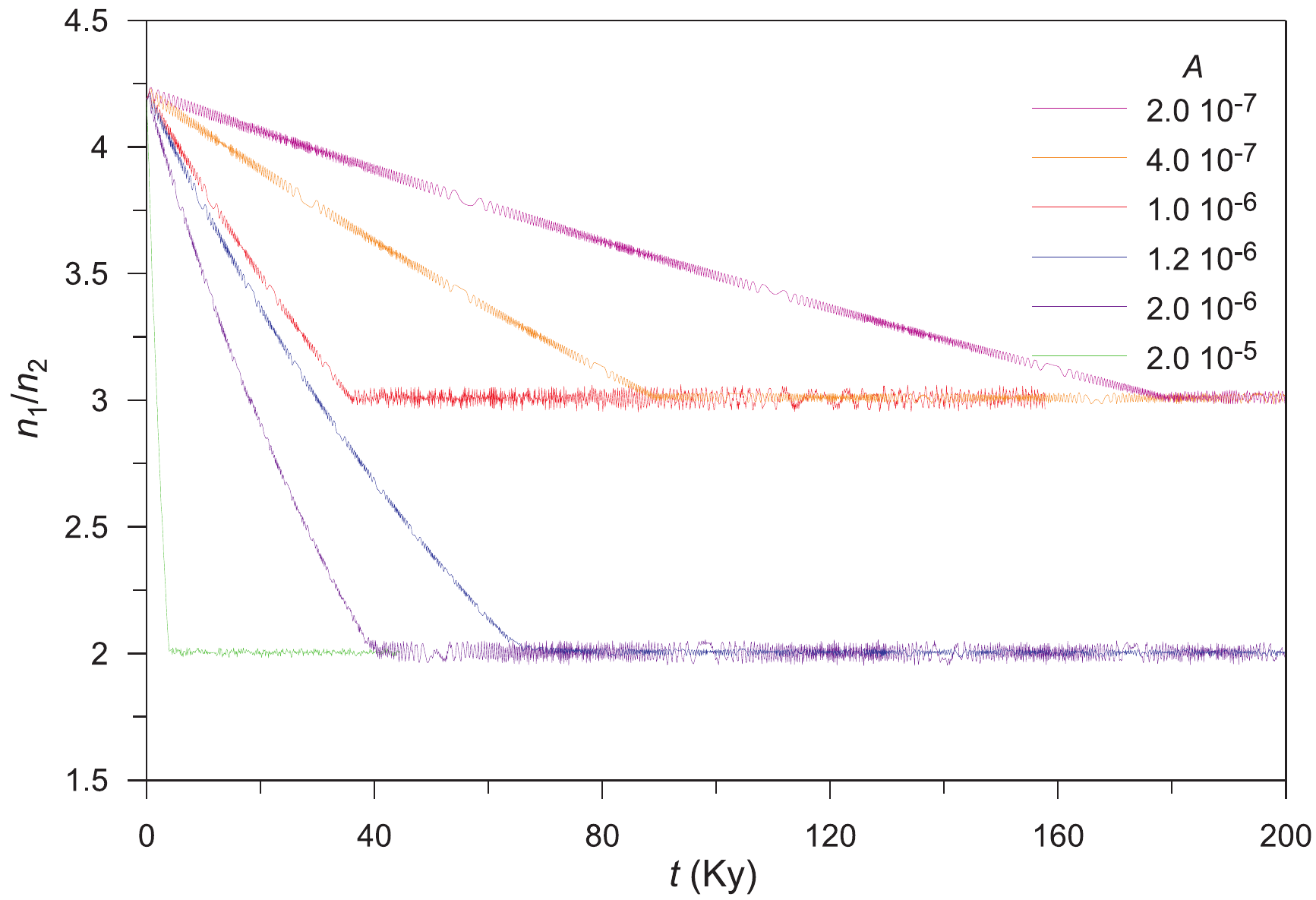}} 
\caption{Resonance capture after planetary migration for different damping rate values $A_2$. There is a critical value, $10^{-6}<A_2^*<1.2\,10^{-6}$, above or below to it the system is captured in the resonance 2:1 or 3:1, respectively.}
\label{figCCupt}       
\end{figure}

Apart from the condition (\ref{EqRCcon}), which is necessary for resonance capture, the decreasing rate of the mean motion ratio $n_1/n_2$ plays also an important role. According to the setup of simulations used above (where we have set $C_1=0$), this rate depends on the parameter $A_2=2C_2(1-a_2)$, which estimates the damping of the semimajor axis, $a_2$, of the outer planet. In Fig. \ref{figCCupt} we present the evolution of $n_1/n_2$ of a system of two planets with $m_1=m_2=0.001$ by setting $\alpha_2=0.9$ in all cases but different values for parameter the $C_2$ (or $A_2$, equivalently). The system starts at the value $n_1/n_2=4.2$ and initially decreases along the evolution of the planetary orbits. For $C_2\leq 5\,10^{-6}$ the system is captured in the 3:1 resonance. When the decreasing rate becomes sufficiently fast ($C_2\geq 6\,10^{-6}$) the system passes through the 3:1 resonance without capture. In these cases the system approaches the 2:1 resonance and is trapped there. 

It is worthy to note that the system is neither captured in the 4:1 nor the 5:2 resonance. This situation is also the case in many other simulations where capture in 2:1 and 3:1 resonances is the most probable \cite{Sylvio03}. Capture in the $4:1$ resonance may be possible for very slow damping rates. On the other hand, the capture in the 5:2 resonance is observed only temporarily \cite{AntoniadouPhD} and for very specific values of the system parameters and initial conditions. In \cite{LibTsi09} some cases of capture in the resonances 4:1, 5:1, 5:2 and 7:2 are reported for higher initial eccentricities but it is not clarified whether such captures are temporal or stable for long time intervals.

\subsection{Migration paths and families of periodic orbits} \label{Section32}
As we mentioned in the introduction, in the general TBP we can compute families of periodic orbits in a rotating frame of reference \cite{Hadjidem75}. These families are classified in two types (i) families of circular periodic orbits and (ii) families of elliptic periodic orbits. Along any family of circular orbits the ratio of semimajor axes $a_2/a_1$ (and equivalently the ratio $n_1/n_2$) varies but the planetary eccentricities are almost zero, $e_i\approx 0$. The circular family has gaps at resonances of the form $\frac{p+1}{p}$, where $p$ is a positive integer. The circular family consist in general of linearly stable periodic orbits except for some sections at the resonances $\frac{n_1}{n_2}=\frac{p+2}{p}$, where $p$ is an odd positive integer.  

\begin{figure}
\centering
\resizebox{0.7\textwidth}{!}{\includegraphics{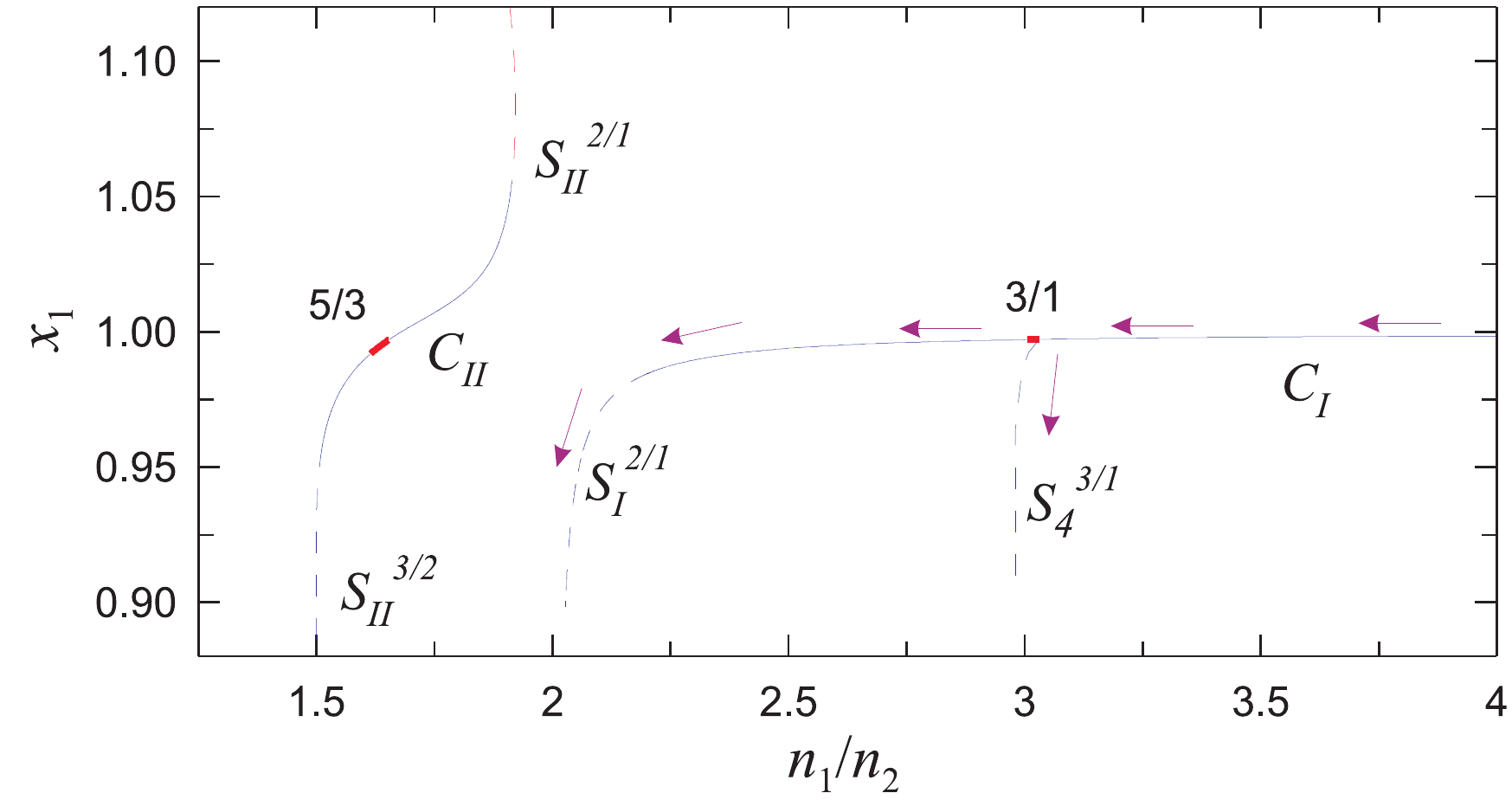}} 
\caption{The elliptic families $S$ and the segments $C_{I}$ and $C_{II}$ of the circular family for planetary masses $m_1=0.001$, $m_2=0.002$. The families are projected in the plane $n_1/n_2$ - $x_1$, where $x_1$ indicates the initial distance of the inner planet $P_1$ from the center of mass $P_1$-star. Blue (red) sections indicate stable (unstable) orbits. The dashed lines indicate the elliptic families $S$.}
\label{figCFam}       
\end{figure}

In Fig. \ref{figCFam}, we present the segments $C_I$ and $C_{II}$ of the circular family for planetary masses $m_1=0.001$, $m_2=0.002$, which are separated with a gap at the 2:1 resonance.  At this resonance, the circular segments continue smoothly as families of elliptic orbits, $S_I^{2/1}$ and $S_{II}^{2/1}$ , which extent up to high eccentricities. Along elliptic families the mean motion ratio $n_1/n_2$ remains almost constant and, therefore, elliptic families are called also {\em resonant families}. At the 3:1 resonance the circular family exhibits an unstable segment but the stable family $S_4^{3/1}$ bifurcates from there. This family is elliptic and along it we have $n_1/n_2\approx 3$ and the eccentricities increase \cite{GV06}.   

In the simulations of the previous subsection, we start with almost zero eccentricities i.e. we are close to an orbit of the circular family. As the system migrates, the ratio $n_1/n_2$ decreases, while the eccentricities remain close to zero (see the arrows in Fig. \ref{figCFam}). Thus, the system migrates along the circular family \cite{Hadjidem10,Hadjidem11}. When the system reaches the 3:1 resonance there are two possibilities: either continue along the circular family or follow the resonant family $S_4^{3/1}$. In the second case, we obtain trapping in the 3:1 resonance and increase of the eccentricities. In the first case, the system continues towards the 2:1 resonance and, due to the gap there, the only possibility for the system is to follow the resonant family $S_I^{2/1}$. As we have mentioned above, the case which is selected by the system depends on the migration rate. At the resonances between the 2:1 and 3:1 (e.g. the 5:2) there are also bifurcations of resonant families without the presence of instabilities. These bifurcating families consist of orbits of higher multiplicity \cite{Psychoyos05}. These properties of the circular family explain why the system is hardly captured in such resonances.   

When the system leaves the circular family, it follows the elliptic - resonant families along which the resonance remains almost constant. These families are symmetric, $S$, i.e. in the rotating frame they are symmetric with respect to the $x$-axis (see Fig. \ref{Fig_Porbs}a,b). In the inertial plane of motion, symmetric orbits correspond to almost Keplerian orbits which are aligned ($\varpi_2-\varpi_1=0$) or anti-aligned ($\varpi_2-\varpi_1=180^\circ$) and the planets are initially (and periodically) in conjunction, i.e. $M_i=0$ or $180^\circ$. From these symmetric families asymmetric families, $A$, may bifurcate \cite{Hadjidem06b}. At the bifurcation points the symmetric families become unstable, while the bifurcated asymmetric families are stable. For particular mass values the families form characteristic curves in phase space. The periodic orbits in these families correspond to elliptic orbits with eccentricities that vary along the families and we can present them as curves on the plane $e_1-e_2$. The form of these curves depends on the mass ratio $m_1/m_2$ (up to first order in masses) \cite{Beauge05}.           

\begin{figure}
\centering
$\begin{array}{ccc}
\resizebox{0.45\textwidth}{!}{\includegraphics{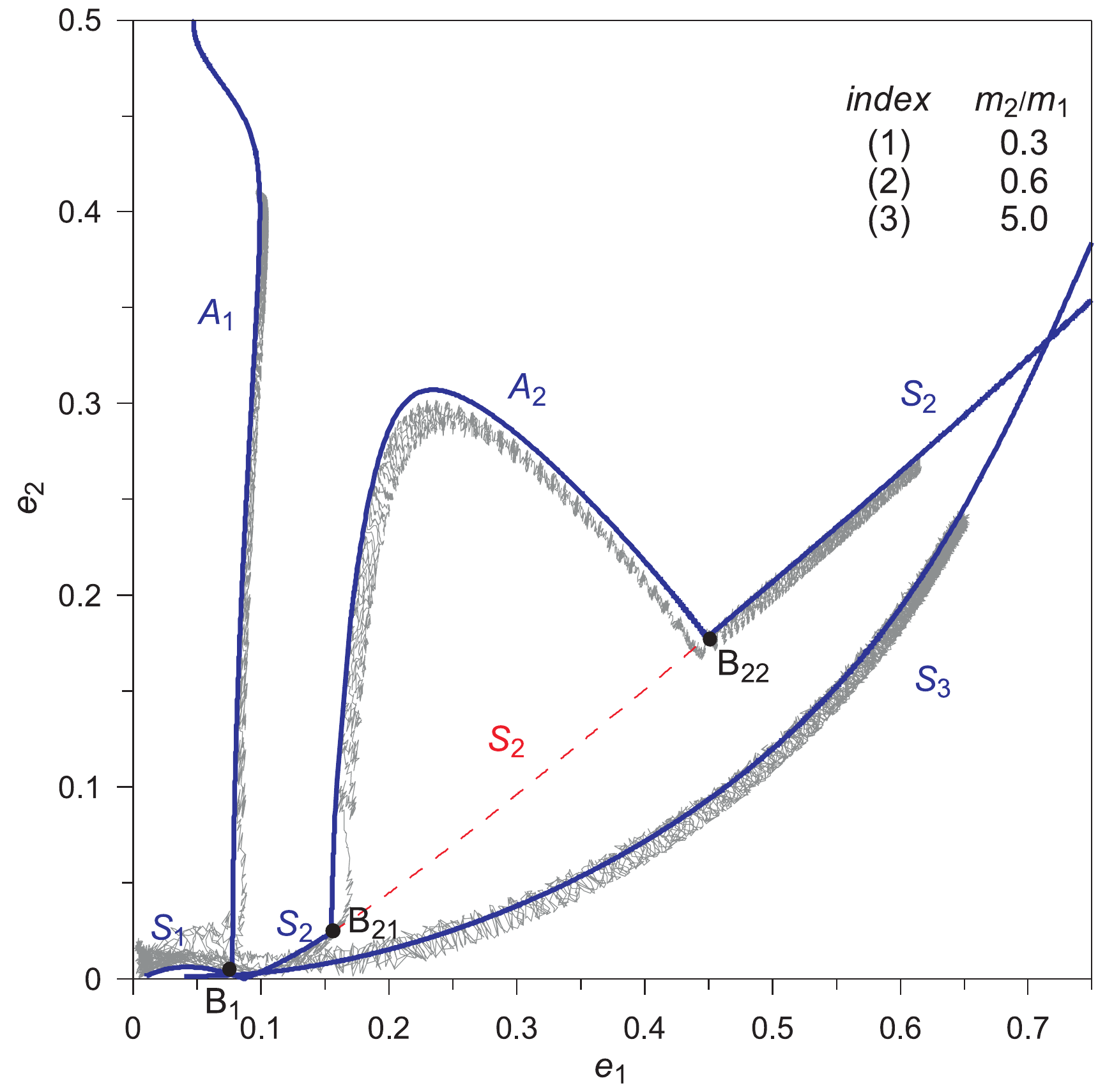}}&\;\;\;\;&\resizebox{0.45\textwidth}{!}{\includegraphics{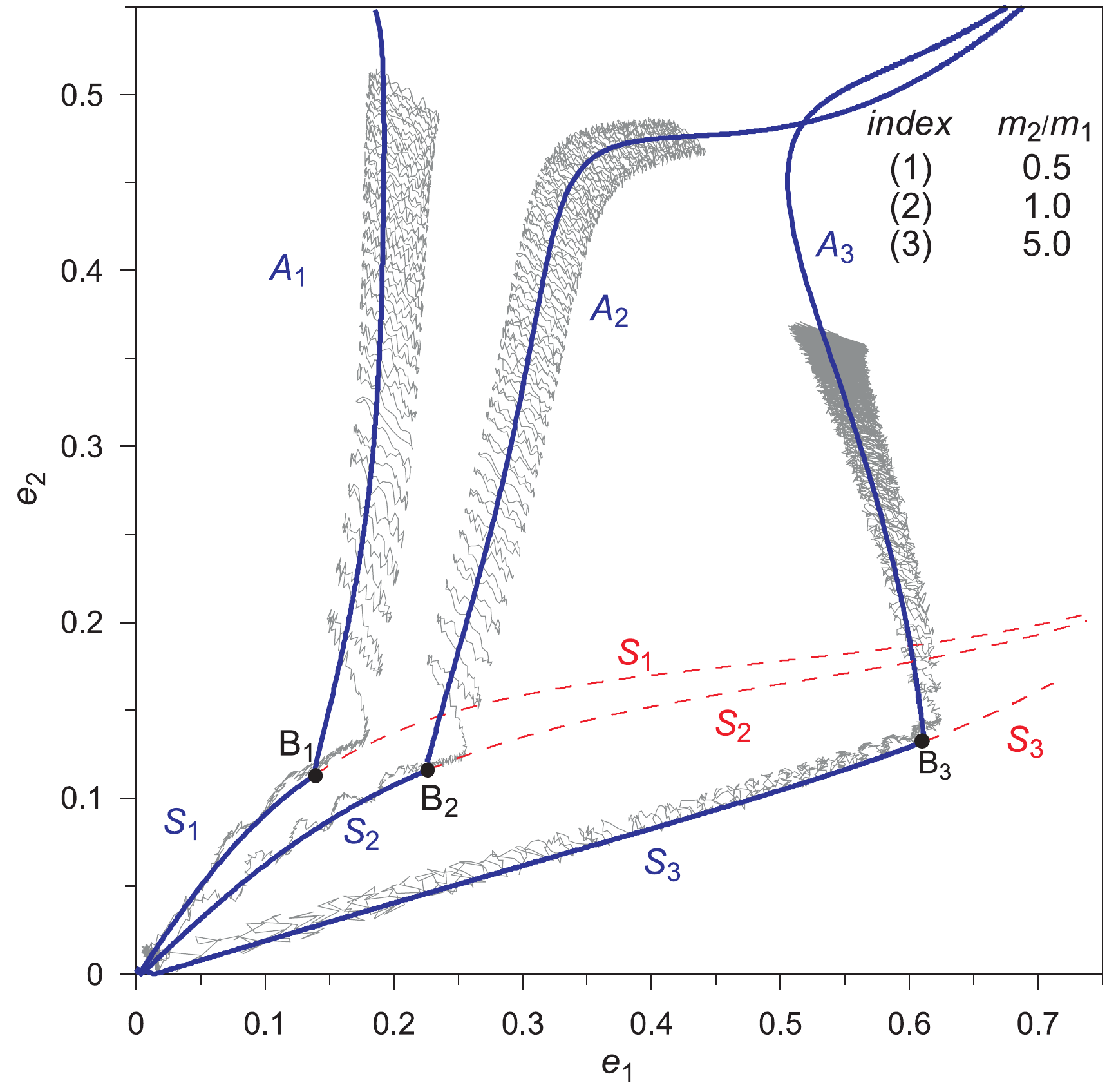}}\\
\textnormal{(a)}&\;\;\;\; & \textnormal{(b)}\\
\end{array}
$
\caption{Families of (a) 2:1  and (b) 3:1 resonant periodic orbits for the indicated mass ratios. Families $S_i$ are symmetric and $A_i$ are asymmetric, which bifurcate from symmetric families at the points $B_i$. Blue (red) sections indicate stable (unstable) orbits. Gray lines show the migrating evolution of the system to higher eccentricities after its trapping in the resonance at about $(0,0)$.}   
\label{fig2131Fams}       
\end{figure}   
   
If Fig. \ref{fig2131Fams}, we present the 2:1 and 3:1 resonant families for the indicated mass ratio values (see \cite{GV09} and \cite{GV08}, respectively). The families bifurcate from the circular family at $(e_1,e_2)=(0,0)$ and are symmetric and stable. At the points $B_i$ the families become unstable and bifurcation of asymmetric families occurs. Particularly, for the mass ratio $m_2/m_1=0.6$ the symmetric family $S_2$ becomes unstable at $B_{21}$ and stable, again, at $B_{22}$. The asymmetric family $A_2$ forms a bridge between the two points. In the same figures, we present the migrating evolution of the planetary system under the Stoke's dissipation with $C_2=10^{-6}$ and $\alpha_2=0.7$ (2:1 case) and $0.6$ (3:1 case). Initially the system evolves along the circular family at (0,0). When the system is captured in the resonance then the stable resonant families are followed. It is clear that the families of stable periodic orbits guide the migration of the planetary system. The migration stops asymptotically at a point ($e_1^*, e_2^*$), which depends on the eccentricity damping parameter $\alpha$. Then the system oscillates around the stable periodic orbit at ($e_1^*, e_2^*$) and, consequently, the migration ends in a stable resonant planetary configuration.             	

\begin{figure}
\centering
\resizebox{0.85\textwidth}{!}{\includegraphics{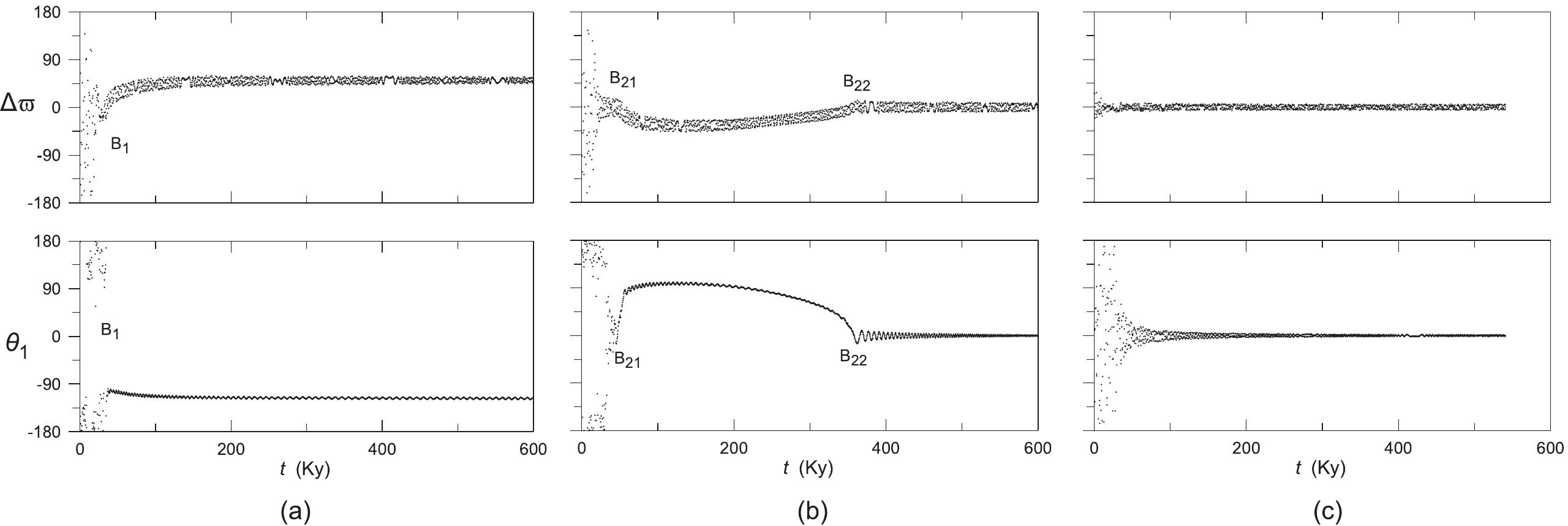}} 
\caption{Evolution of resonant angles along the 2:1 resonant families for planetary mass ratio (a) $m_2/m_1=0.3$ (b) $m_2/m_1=0.6$  (c) $m_2/m_1=5.0$}
\label{figRang21}       
\end{figure}
	
In a $p:q$ ($p\neq q$) resonance we can define the resonant (slow) angle variables \cite{Morbi02}
\begin{equation}
\theta_i=p\lambda_2-q\lambda_1-(p-q)\varpi_i,\;\;(i=1,2),  \quad\quad \Delta\varpi=\varpi_2-\varpi_1,
\label{EqResAngle}
\end{equation}
where $\lambda_i=M_i+\varpi_i$ is the mean longitude of the planet $P_i$. The apsidal difference $\Delta\varpi$ is easily derived from $\theta_i$. At the exact resonance or, equivalently,  at a periodic orbit, $\theta_i$ are almost constant and this is also called {\em apsidal corotation resonance}. For elliptic symmetric periodic orbits $\theta_i=0^\circ$ or $180^o$ and this value characterizes all the periodic orbits in family sections where $e_i\neq0$. Instead, in asymmetric periodic orbits $\theta_i$ can take any value which varies along the family. In Fig. \ref{figRang21} we present the evolution of $\theta_1$ and $\Delta\varpi$ as the system migrates along the 2:1 resonant families. After resonance capture the angles librate with small amplitude around the value that corresponds to the exact resonance.
E.g. in the panel (c) the angle $\theta_1$ initially rotates but the apsidal difference $\Delta\varpi$ librates even before the capture (this is called {\em apsidal resonance}). After the capture in the exact resonance both angles librate around $0^\circ$ (symmetric configuration). In the panels (a) and (b) we observe the asymmetric planetary configurations after $B_1$ and between $B_{21}$ and $B_{22}$, respectively.


\section{Migration of three-planet systems}    
The dynamics of three-planet systems is described by the four-body problem of planetary type for which only few works can be found in literature. The system of Jupiter's satellites Io - Europa - Ganymede has been studied firstly by Ferraz-Mello \cite{Ferraz1979}. This system is locked in the {\em Laplace resonance} i.e. Io-Europa revolves in the 2:1 resonance and the same holds for the Europa-Ganymedes pair. Also, computation of periodic orbits for this system are found in \cite{HadjiMich81}. The Laplace resonance of the exosolar system Gliese-876 is studied in \cite{Marti13}. Numerical simulations for the study of resonance trapping have been performed by Libert and Tsiganis \cite{LibTsi11}. They used the system (\ref{EqDeqs}) by applying the Stoke's drag to the outer and the middle planet. They indicated the particular resonances observed and the possible increase of inclinations but they do not examine the long term stability of the resonant configurations obtained. 

In the following we present particular numerical simulations which lead to resonance capture and, finally, to a stable planetary configuration. In our approach, we set the inner ($P_1$) and the middle planet ($P_2$) in a regular eccentric orbit at the 2:1 resonance, as those described in the previous section (i.e. close to a stable periodic orbit). We assume that the dissipative force is not applied to these planets. The third planet ($P_3$) is assumed to migrate due to the Stoke's force starting at a large distance from the orbit of $P_2$. 

\begin{figure}
\centering
\resizebox{0.85\textwidth}{!}{\includegraphics{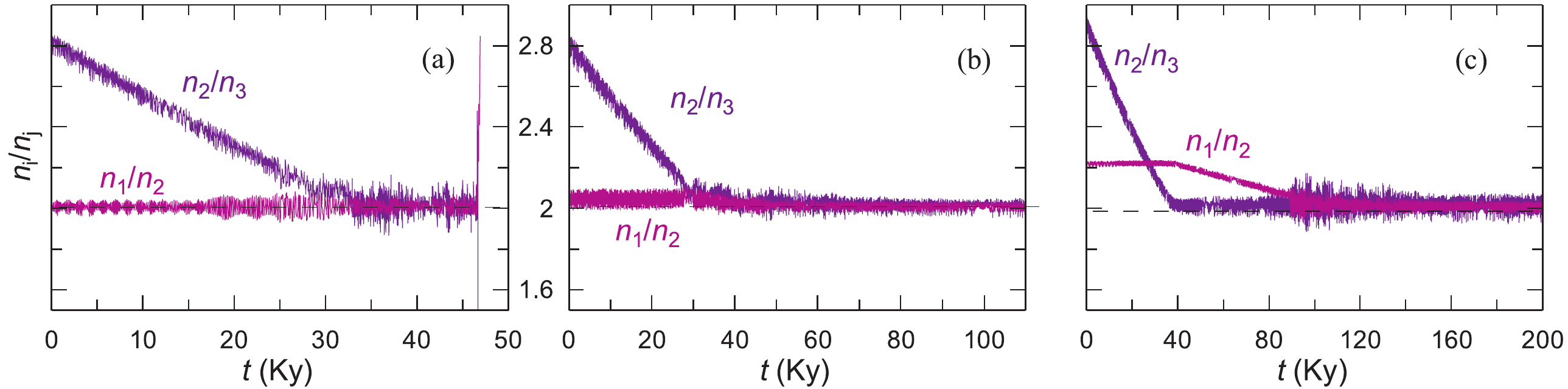}} 
\caption{Evolution of the mean motion ratio of the planet pairs $P_1-P_2$ and $P_2-P3$ during migration. The planetary masses are $m_1=0.001$, $m_2=0.002$, $m_3=0.005$ and $C_3=10^{-5}$, $\alpha=0.95$. $P_3$ starts from a distant circular orbit  (a) The planets $P_1$, $P_2$ are initially in eccentric 2:1 resonant motion    (b) $P_1$, $P_2$ are initially in 2:1 resonant circular orbits  (c) $P_1$, $P_2$ are initially at circular but non-resonant orbits.}
\label{fignn21}       
\end{figure}

The simulations presented in Fig. \ref{fignn21} are quite typical for planetary masses of the order of Jupiter's mass (i.e. $m_i=O(10^{-3})$). In panel (a) $P_1$ and $P_2$ are located in a configuration, which corresponds to a stable symmetric elliptic periodic orbit in the 2:1 resonance ($a_1=1$, $a_2=1.58$) and at the eccentricities $e_1=0.4$, $e_2=0.11$. $P_3$ starts from an almost circular orbit at $a_3=3.2$ (i.e. $n_2/n_3\approx 2.8$) and migrates inward due to the drag force with $C_3=10^{-5}$ and $\alpha=0.95$, i.e. $n_2/n_3$ decreases. The capture in the three-planet resonance 1:2:4, which is known as {\em Laplace resonance}, is obtained after about $35Ky$ when $n_2/n_3\approx 2$. However, the trapping in the resonance holds only for about $10Ky$ and then the system leaves the resonance.  We mention that the eccentricities show strong irregular variations after the resonance capture and the motion becomes strongly chaotic. Under such conditions close encounters are unavoidable and the system is disrupted. Therefore, in this case the capture is temporal and the particular planetary configuration during the capture is characterized as unstable. Many numerical simulations show that if the inner planet pair starts from an eccentric motion, the capture in a stable three-planet configuration is quite rare.  

In the second simulation (Fig. \ref{fignn21}b) we consider the same setup as above but now the inner and the outer planet starts from an almost circular orbit $e_i\approx 0$ at the 2:1 resonance. During migration the planets $P_1$ and $P_2$ remain in 2:1 resonance. When $n_2/n_3\approx 2$, $P_3$ is also trapped in the 2:1 resonance with $P_2$. Thus, we obtain trapping in the Laplace resonance which lasts for a very long time interval. If the planets $P_1$ and $P_2$ are initially not in resonance but $n_1/n_2$ is a bit larger than $2.0$, then we obtain the evolution shown in Fig. \ref{fignn21}c. The mean motion ratio $n_2/n_3$ initially decreases due to the inward migration of $P_3$ but $n_1/n_2$ remains constant in average. When $n_2/n_3\approx 2$, $P_2$ and $P_3$ are captured in the resonance and this causes a decreasing rate for $n_1/n_2$. When $n_1/n_2$ reaches the 2:1 resonance, too, we get a stable 1:2:4 resonant configuration and the rest evolution is qualitatively the same as in the previous simulation (panel b). The above results have been verified by many simulations, i.e. in order to create a stable resonant configuration we should start migration from orbits with low eccentricities.  

For three-planet systems we can define resonant angles, similarly to the two-planet systems (see Eq. (\ref{EqResAngle})). For the Laplace resonance these angles are defined as \cite{Marti13}
\begin{equation}
\begin{array}{ccc}
\theta_1=\lambda_1-2\lambda_2+\varpi_1 & \quad & \theta_2=\lambda_1-2\lambda_2+\varpi_2 \\
\theta_3=\lambda_2-2\lambda_3+\varpi_2 & \quad & \theta_4=\lambda_2-2\lambda_3+\varpi_3
\end{array}
\label{eqrth4}
\end{equation}
Note that in the above definition we take separately the angles of the two body resonances. In the neigbourhood of the exact resonance all resonant angles should librate. We can easily obtain from Eq. (\ref{eqrth4}) the relations
\begin{equation}
\Delta \varpi_{12}=\varpi_2-\varpi_1=\theta_2-\theta_1, \quad \Delta \varpi_{23}=\varpi_3-\varpi_2=\theta_4-\theta_3.
\label{eqDw}
\end{equation}
Thus, in the resonance the apsidal differences of the separate pairs of planets should librate too. Also we can derive the {\em Laplace resonant angle}, which depends only on the mean longitudes $\lambda_i=M_i+\varpi_i$ and is defined as
\begin{equation}
\theta_L=\theta_2-\theta_3=\lambda_1-3\lambda_2+2\lambda_3.
\label{eqThL}
\end{equation}
The libration of $\theta_L$ is a strong indication that the system has been trapped in the resonance.

\begin{figure}
\centering
$\begin{array}{ccc}
\resizebox{0.42\textwidth}{!}{\includegraphics{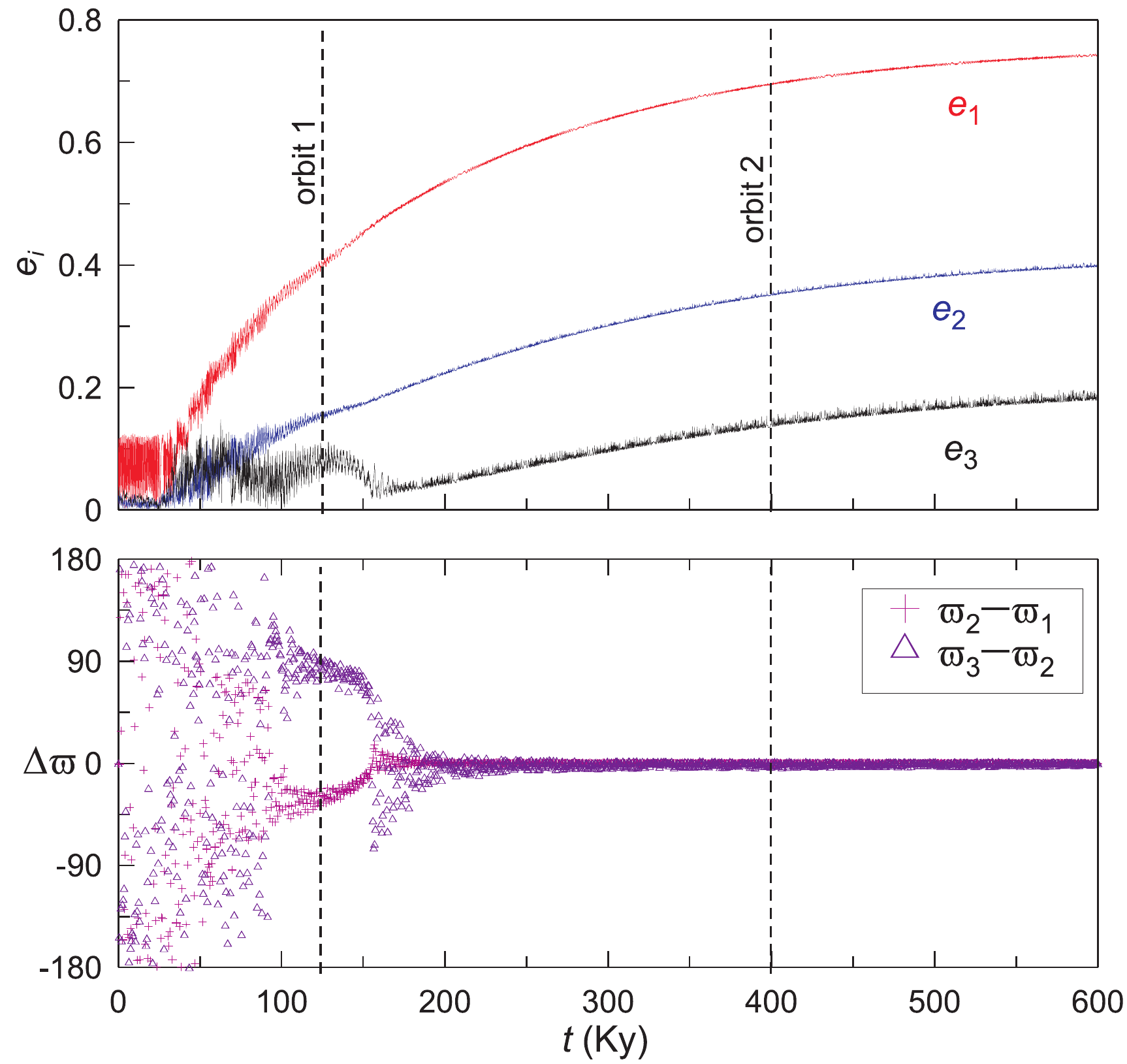}}&\;\;\;&\resizebox{0.42\textwidth}{!}{\includegraphics{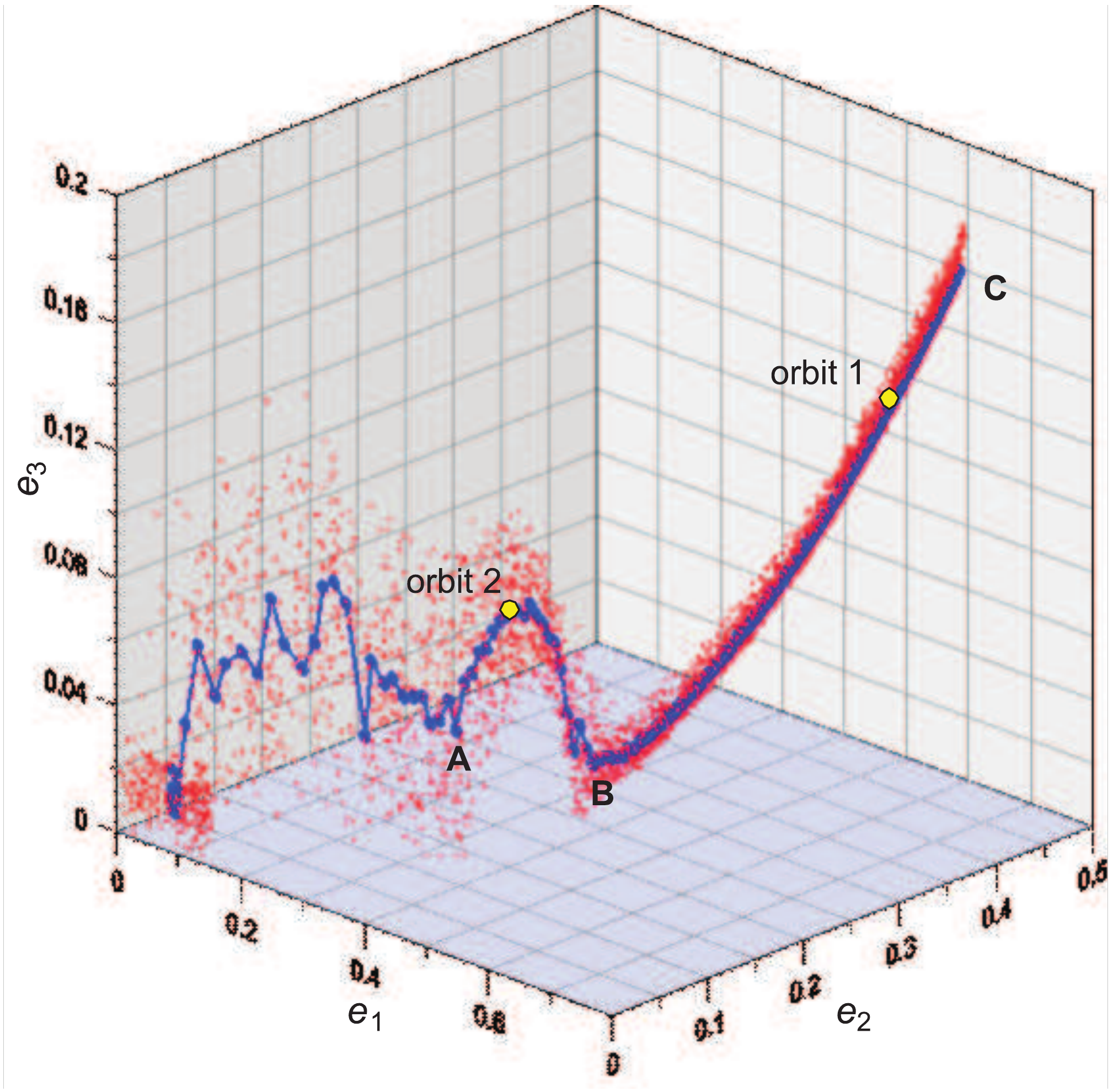}}\\
\textnormal{(a)}&\;\;\; & \textnormal{(b)}\\
\end{array}
$
\caption{(a) Evolution of eccentricities and apsidal angles $\Delta\varpi_{ij}$ during the migration presented in Fig. \ref{fignn21}b. (b) the presentation of the migration in the eccentricity space $e_1-e_2-e_3$ (red crosses). Blue line indicates the running average of the evolution (the migration path). A symmetric and an asymmetric orbital stage at $t=400Ky$ ({\em orbit 1}) and at $t=125Ky$ ({\em orbit 2}), respectively, are indicated.}   
\label{fig124mig}       
\end{figure}   

Next we focus our study on the stable resonant capture presented in Fig. \ref{fignn21}b.  In Fig. \ref{fig124mig}a, we show the evolution of the eccentricities and the apsidal difference angles along the simulation. We observe three intervals of different behaviour. Before capture, the eccentricities show small oscillations at small values (larger amplitude is observed for $e_1$). At this interval $\Delta\varpi_{12}$ and $\Delta\varpi_{23}$ rotate. After the resonance capture, at $t\approx 25Ky$, the eccentricities $e_1$ and $e_2$ start to increase while $e_3$ shows some fast and slow variations of a relatively large amplitude. At this interval the angles $\Delta\varpi_{ij}$ seem to librate around asymmetric values (i.e. different than $0^\circ$ or $180^\circ$). However, this situation is temporal and for $t>200Ky$ we get librations around $0^\circ$. Also for $t>150Ky$, $e_3$ starts to increase quite smoothly. Thus the system can be clearly located at particular orbital elements. Similarly with Fig. \ref{fig2131Fams}, if we present the evolution in the space of planetary eccentricities we obtain the {\em migration path} of Fig.\ref{fig124mig}b. After the initial large variations at low eccentricities the system seems to follow a particular characteristic curve. In the arc AB there are asymmetric librations of the resonant angles while at section BC all resonant angles librate around $0^\circ$. Accordingly with the systems of two planets, we may conjecture that the migration path reveals families of stable periodic orbits.     

In order to determine initial conditions, $a_{i0}$, $e_{i0}$, $\varpi_{i0}$, $M_{i0}$ ($i=1,2,3$), close to the exact resonance, we perform a numerical integration by considering as initial conditions those provided by the migration evolution at particular time values where the system seems to have reached a stable configuration. After that point we stop including the dissipative force in the integration and allow, instead, the system to evolve only with gravitational interactions.  If the system is indeed close to the exact resonance we expect that $a_i(t)\approx a_{0i}$ and $e_i(t)\approx e_{i0}$. We determine the values $a_{0i}$ and $e_{0i}$ as the averages $<a_i(t)>$ and $<e_i(t)>$, respectively, in the interval of integration (of about $1Ky$). The pericenter angles $\varpi_{0i}$ can be obtained from Eqs. (\ref{eqDw}) by taking into account the libration centers of $\Delta\varpi_{12}$ and $\Delta\varpi_{23}$, which are determined again by averages over the integration interval and by choosing an inertial reference frame where $\varpi_1=0$. In the same way, by using Eqs. (\ref{eqrth4}),  we can determine the location of the planets on their elliptic orbits, e.g. the initial mean anomalies $M_{0i}$, from the libration centers of the resonance angles $\theta_2$ and $\theta_3$ by assuming that the inner planet is initially at its pericenter ($M_{10}=0$). 

\begin{figure}
\centering
$\begin{array}{ccc}
\resizebox{0.42\columnwidth}{!}{\includegraphics{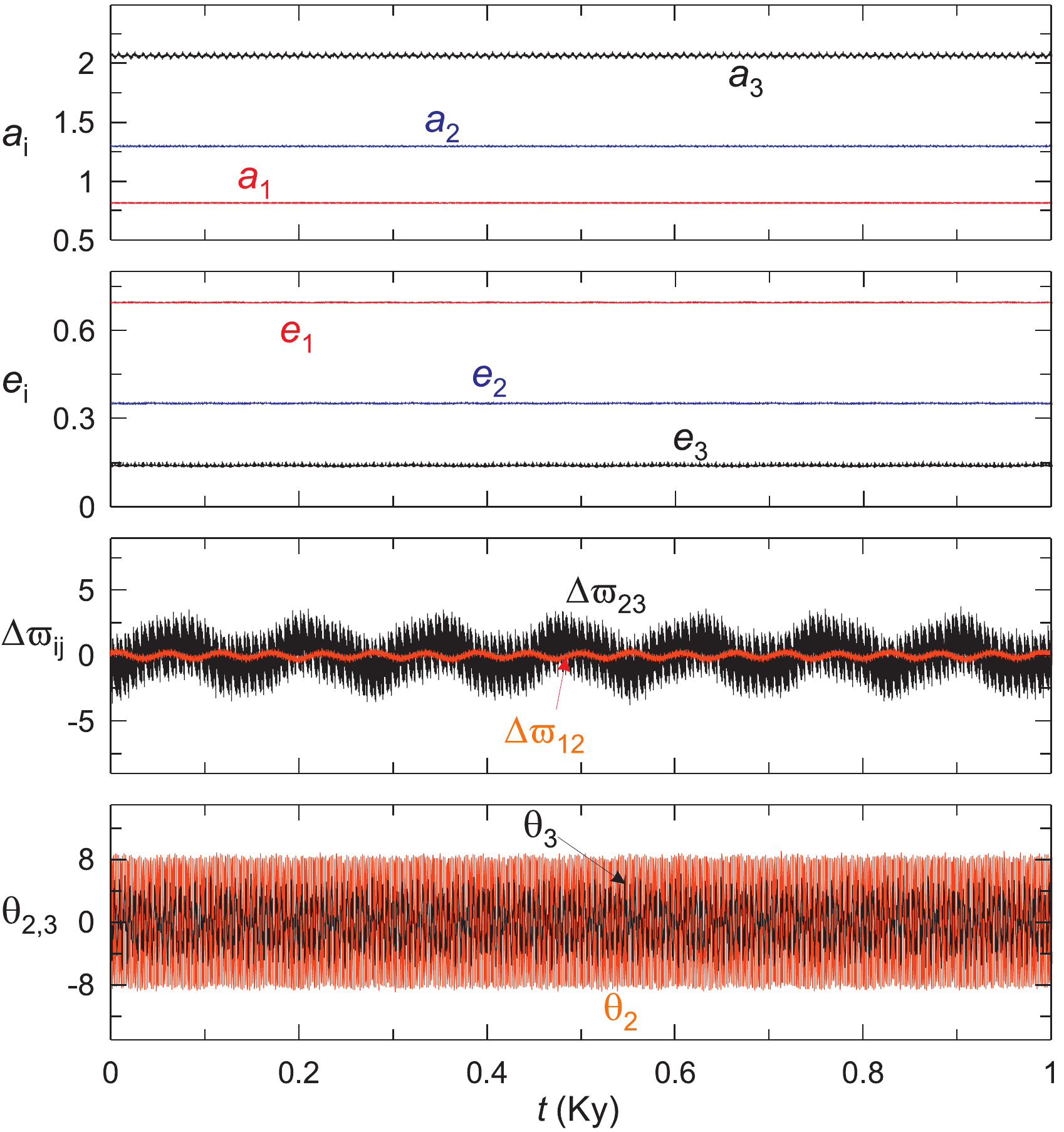}}&\;\;\;&\resizebox{0.42\columnwidth}{!}{\includegraphics{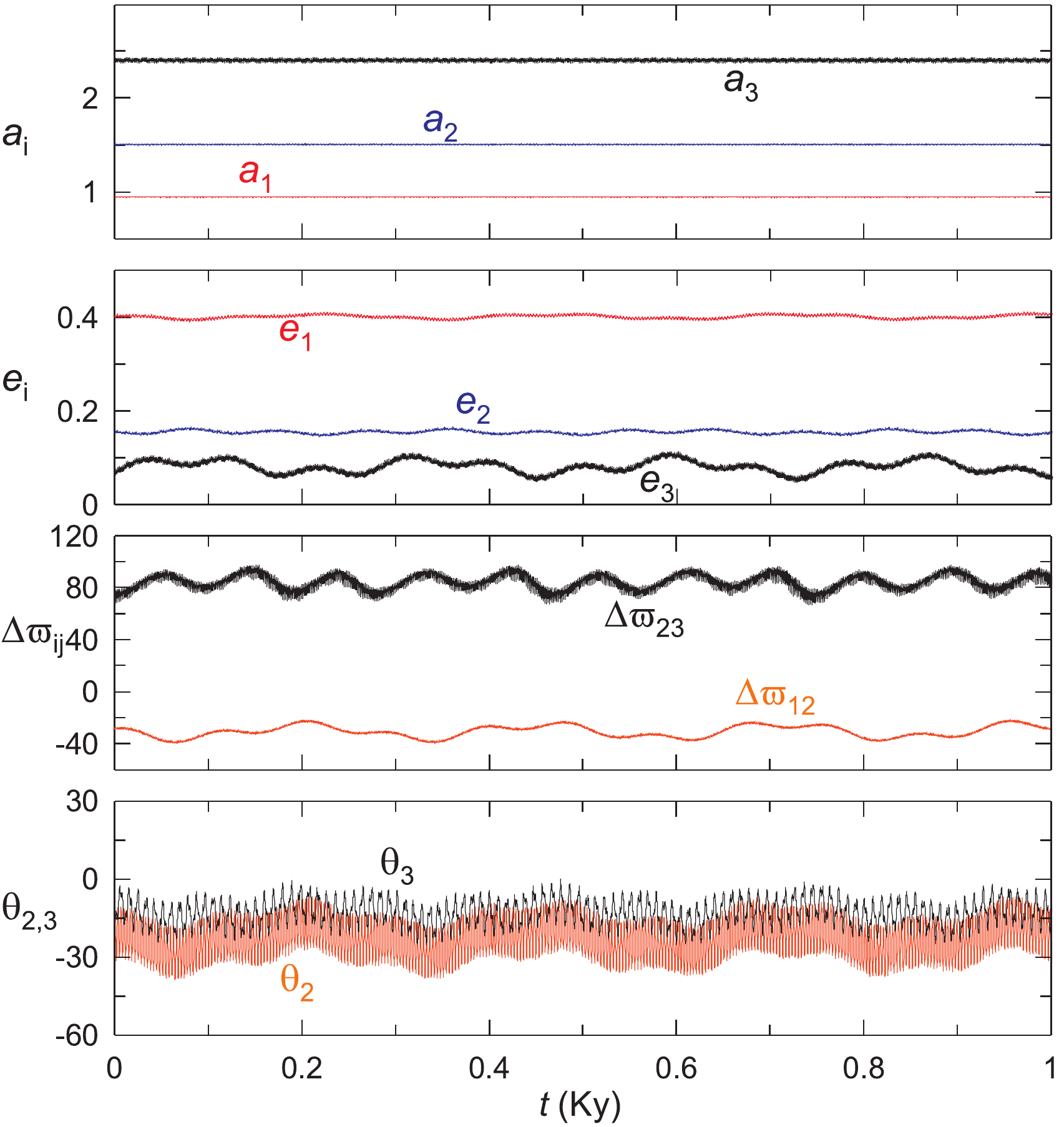}}\\
\textnormal{(a)}&\;\;\; & \textnormal{(b)}\\
\end{array}
$
\caption{Evolution of orbital elements and resonant angles for the (a) symmetric $orbit\;1$  and (b) asymmetric $orbit\;2$.}   
\label{figEvolsas}       
\end{figure}   

\begin{figure}
\centering
\resizebox{0.85\columnwidth}{!}{\includegraphics{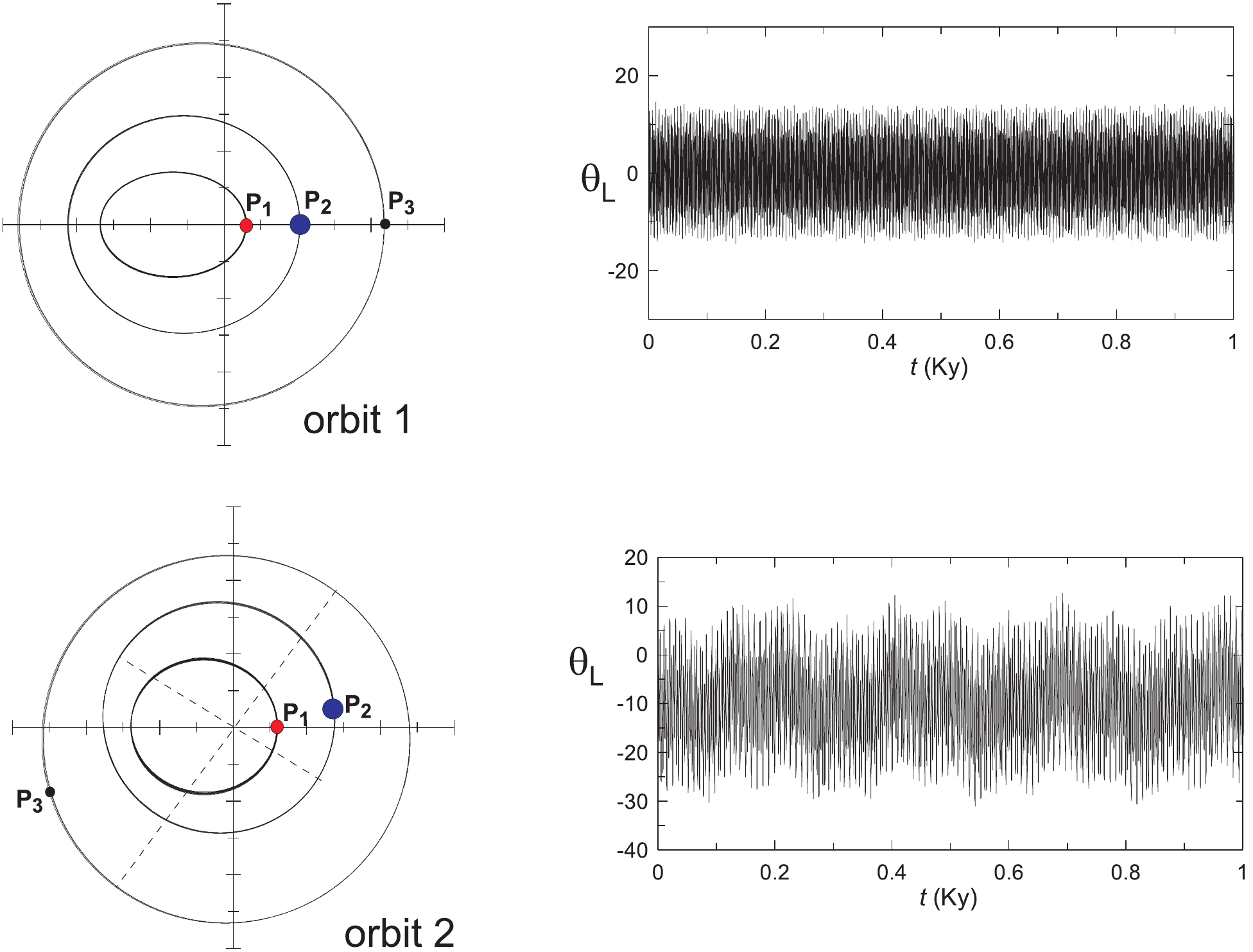}} 
\caption{The Keplerian planetary orbits for the stable configurations of {\em orbit 1} and {\em orbit 2} and the evolution of the Laplace resonant angle $\theta_L$. }
\label{figOrb}       
\end{figure}

As an example we present in Fig. \ref{figEvolsas} the evolution of some orbital elements for two orbits with initial conditions taken from the migration evolution at $t=400Ky$ ({\em orbit 1}) and at $t=125Ky$ ({\em orbit 2}), as they are indicated in Fig. \ref{fig124mig}. Applying the method described above, we take the following initial conditions of stable configurations:
$$
\begin{array}{lllll}
Orbit\;1: & a_{10}=1.0,& e_{10}=0.695, & \varpi_{10}=0^\circ, &  M_{10}=0^\circ\\
& a_{20}=1.589, & e_{20}=0.341, & \varpi_{20}=0^\circ, & M_{20}=0^\circ\\
& a_{30}=2.528, & e_{30}=0.139, & \varpi_{30}=0^\circ, & M_{30}=0^\circ \\
\end{array}
$$
and 
$$
\begin{array}{lllll}
Orbit\;2: & a_{10}=1.0, & e_{10}=0.400, & \varpi_{10}=0^\circ, & M_{10}=0^\circ \\
& a_{20}=1.591, & e_{20}=0.154, & \varpi_{20}=-30.6^\circ, & M_{20}=26.6^\circ\\
& a_{30}=2.536, & e_{30}=0.081, & \varpi_{30}=53.7^\circ, & M_{30}=-64.5^\circ
\end{array}
$$
As expected, {\em orbit 1} is symmetric, particularly all planetary ellipses are aligned and planets can be set initially at their pericenters. {\em Orbit 2} is asymmetric and lines of apsides of planetary orbits form angles different than $0^\circ$ or $180^\circ$. Also, the planets cannot be found in conjunction i.e. they can not be found at the same time at an apside (pericenter or apocenter). The planetary orbits (for a short time interval) and a potential initial position of planets is presented in Fig. \ref{figOrb}.  Also we present the libration of the Laplace resonant angle (\ref{eqThL}) along these orbits, which indicates that the system is inside the 1:2:4 resonance.        
    
\section{Conclusions}
We studied the possible resonance capture of planets and its consequences in the evolution of  planetary systems. Starting from a planetary system, which, generally, is not formed within a resonant domain, we consider a migration process that drives the planets in a resonant trapping. From a physical point of view we can assume that planetary migration is caused by the interaction of the planets with the proto-planetary disk of gas and dust. In our study we use the Stoke's drag force to simulate this interaction. The effect of this additional force has as a consequence the damping of semimajor axes and the damping of eccentricities of planetary orbits.

Our main aim of this study is to determine, due to resonant capture, planetary configurations that guarantee  long-term stability.  We showed that resonant capture is probable when particular initial conditions are assumed and the parameters of the dissipative forces are restricted in some intervals. The planetary configurations, which are obtained after migration, may be quite eccentric, since capture in resonance causes a rise in the eccentricities when eccentricity damping is not very strong. 

In section 3, we reviewed the dynamics of 2:1 and 3:1 resonant capture which is closely related with the periodic orbits of the three-body problem. Such periodic orbits are non isolated but they form families of orbits, which mono-parametrically extend in phase space and along the particular resonance. When the system is captured in the resonance then its migration continues due to the drag and the average evolution in phase space is driven by the families of stable periodic orbits. In the present study we give a possible explanation why the capture in the 2:1 resonance is very probable and consists a normal limit of the migration process. Capture in 3:1 resonance is also probable due to the existence of a small unstable segment in the circular family and the bifurcation of a stable family of periodic orbits. However, in this case the migration rate should be sufficiently slow. On the other hand, and according to our approach, the permanent capture in the 5:2 resonance is not so probable. Since the families of periodic orbits depends on the ratio of planetary masses, the same is true for the migration paths of the planetary systems. This becomes clear in Fig. \ref{fig2131Fams} and we may claim that the same dependence also holds for three-planet systems. 

The dynamics of three-planet systems is more complicated and has not been studied sufficiently yet. Various simulations show that most of such systems are unstable and disrupted in short time intervals. A first numerical study which considers the effect of dissipative forces has been done in \cite{LibTsi11}. In the present paper, we based on the dynamics of two-planet systems and proposed particular initial and migration conditions in order to obtain trapping in stable three-planet resonances. It seems that, when we consider slow migration rate and almost circular initial planetary orbits, the system is driven in a three-planet resonant stable configuration. We studied and showed a typical simulation with capture in the Laplace resonance 1:2:4. Particular planetary configurations of long-term stability can be determined by considering local averages along the migration process and averages of the orbital elements along evolution in the absence of the dissipative force. Capture and stable configurations can be determined also for other resonances, e.g. 1:2:6, 1:3:6 and 1:3:9, by following the same procedure. 

Similarly to the two-planet systems, it seems that migration of three planet systems follows particular paths in phase space that should be related to the existence of families of periodic orbits. The only known relevant computed families are those in \cite{HadjiMich81}. They are symmetric, restricted in low eccentricities and computed for the particular system of the Galilean satellites Io, Europa and Ganymede. In our study, we have found strong indications that resonant families exist and extend up to high eccentricities in the four-body problem of planetary type. Also, these families are either symmetric or asymmetric. The computation of the associated families, which should be done in a future work, will prove definitely the stability of the planetary configurations obtained, when dissipation vanishes. This could provide useful information for understanding the formation and the orbital configuration of multiple exoplanet systems. Also, a similar study can be performed for systems of more that three planets. The determination of stable configurations in such systems is an open question.
      
\vskip 0.5cm
\noindent
{\bf Acknowledgements:} The author would like to thank the MSc student O. Kotsas who managed a large amount of numerical simulations. This research has been co-financed by the European Union (European Social Fund - ESF) and Greek national funds through the Operational Program ``Education and Lifelong Learning'' of the National Strategic Reference Framework (NSRF) - Research Funding Program: Thales. Investing in knowledge society through the European Social Fund.

\end{document}